\documentstyle[12pt]{article}

\newtheorem{theorem}{Theorem}

\newtheorem{theorema}{Theorem~A}
\newtheorem{lemmaa}{Lemma~A}

\begin{document}

\newcommand{\lap}{\bigtriangleup} 
\def\R{{\rm I\kern-.1567em R}}
\def\N{{\rm I\kern-.1567em N}}

\def\ekin{E_{\rm kin}}
\def\epot{E_{\rm pot}}
\def\C{{\cal C}}
\def\F{{\cal F}}
\def\M{{\cal M}}
\def\HC{{\cal H}_C}
\def\proof{{\em Proof.\ }}
\def\prfe{\hspace*{\fill} $\Box$

\smallskip \noindent}

\title{Stable Models of Elliptical Galaxies}
\author{ Yan Guo\\
         Lefschetz Center for Dynamical Systems \\
         Division of Applied Mathematics \\
         Brown University, Providence, RI 02912 \\
         and \\
         Gerhard Rein\\
         Institut f\"ur Mathematik \\ 
         Universit\"at Wien\\
         Strudlhofgasse 4\\
         A-1090 Vienna, Austria}
\date{}
\maketitle

\begin{abstract}
We construct stable axially symmetric models
of elliptical galaxies. The particle density on phase space
for these models depends monotonically on 
the particle energy and on the third component 
of the angular momentum. They are obtained as minimizers of 
suitably defined energy-Casimir functionals, and this implies their
nonlinear stability. Since our analysis proceeds from a rigorous but purely
mathematical point of view it should be 
interesting to determine if any of our models match 
observational data in astrophysics. The main purpose
of these notes is to initiate some exchange of information 
between the astrophysics and the mathematics communities.
\end{abstract}

\section{Introduction}
\setcounter{equation}{0}

Consider a large ensemble of mass points (stars) which interact
only by the gravitational field which they create collectively.
Such a collisionless, self-gravitating gas
is used to model galaxies or globular clusters, 
cf.\ \cite{BT,FP} and the references there. The time evolution of
the density $f = f(t,x,v)\geq 0$ of the stars 
in phase space is governed by 
the following nonlinear system of partial differential
equations:
\[
\partial_t f + v \cdot \partial_x f - \partial_x U \cdot 
\partial_v f = 0 ,
\]
\[ 
\lap U = 4 \pi\, \rho,\ \lim_{|x| \to \infty} U(t,x) = 0 , 
\]
\[
\rho(t,x)= \int f(t,x,v)dv .
\]
Here $t \in \R$ denotes time, $x, v \in \R^3$ denote 
position and velocity respectively, $\rho$ is the spatial mass 
density of the stars, and $U$ the gravitational potential
which the stars induce collectively. Collisional or relativistic
effects are neglected. This system, which
Sir J.~H.~Jeans introduced at the beginning 
of the last century 
for the above modeling purposes, is sometimes in the
astrophysics literature referred to as the collisionless Boltzmann-Poisson
system. Following the mathematics convention we call it the Vlasov-Poisson
system.  

In the present paper we are interested in the steady states of this system.
Besides their very existence, questions of interest are the possible
shapes of such steady states, in particular their symmetry type,
and whether or not they are dynamically stable. 
We approach these questions from a mathematics point of view,
in particular, only such information as can be extracted from
the system stated above is to enter our arguments.
Our ``ideal'' reader is an astrophysicist interested in these
same questions. We find it
deplorable that there is little communication between astrophysicists
and mathematicians investigating these problems, and the present paper is
an attempt to change this. 
Axially symmetric steady states of the the Vlasov-Poisson system were 
obtained in \cite{R5} as perturbations of
spherically symmetric ones via the implicit function theorem.
From an astrophysics point of view axisymmetric models of 
elliptical galaxies have 
been investigated in \cite{pt,t,w}, but as to their stability 
no rigorous results are known to us. 
In the present paper we follow the variational approach developed
in \cite{G1,G2,GR1,GR2,GR3,R3,R6,R4}: We obtain axially symmetric
steady states with vanishing and with non-vanishing velocity field
as minimizers of appropriately defined energy-Casimir functionals.
These steady states are no longer restricted to small
perturbations of spherically symmetric ones, and, most importantly,
they are nonlinearly stable. 
The question we would like to raise is:
Do our mathematical constructions yield 
suitable models for real-world galaxies, or, if not,
how can the mathematical approach be modified to do so? 

If $U_0=U_0(x)$ is a stationary
potential then any function $f_0 = \phi(E)$ of the particle energy
\begin{equation} \label{edef}
E=E(x,v)=\frac{1}{2}|v|^2 + U_0(x)
\end{equation}
solves the Vlasov equation with the potential $U_0$, since 
$E$ is a conserved quantity along particle
trajectories. So if $U_0$
is the potential induced by $f_0$ then this is a steady state
of the Vlasov-Poisson system---at this point it must be emphasized
that to obtain a self-consistent model by this approach, i.~e.,
to make sure that $U_0$ is indeed the potential induced by $f_0$, is 
mathematically non-trivial.
Any steady state obtained by this ansatz
is necessarily spherically symmetric,
as follows from a result of 
Gidas, Ni \& Nirenberg \cite{GNN}. 
In the present paper we make $f_0$
depend on an additional invariant of the particle trajectories.
If $U_0$ is axially symmetric, i.~e., invariant under all rotations
about, say, the $x_3$-axis, then the corresponding component
of angular momentum,
\begin{equation} \label{pdef}
P = P(x,v) = x_1 v_2 - x_2 v_1
\end{equation}
is such an additional invariant. To obtain steady states with a number
density $f_0$ which depends on $E$ and $P$ we minimize the
energy-Casimir functional
\[
\frac{1}{2} \int\!\!\!\int |v|^2 f(x,v)\,dv\,dx
- \frac{1}{2} \int\!\!\!\int \frac{\rho(x)\,\rho(y)}{|x-y|}\,dx\,dy
+ \int\!\!\!\int Q(f(x,v),P(x,v))\,dv\,dx 
\]
subject to the constraint that $f$ is axially symmetric and
has prescribed mass $M>0$:
\[
\int\!\!\!\int f(x,v)\,dv\,dx = M.
\]
The corresponding Euler-Lagrange equation then shows that
any minimizer is a steady state of the desired form.
In order to conclude that this steady state is nonlinearly stable
it is essential that the energy-Casimir functional is conserved
along solutions. This is always true for the sum of the
first two terms which are the kinetic and the potential energy 
of the system. However, the third term is conserved only if
$P$ is constant along particle orbits which is true in general only if 
the solution is axially symmetric. Thus, the stability result which
we derive is restricted to axially symmetric perturbations
of the steady state. The fact that the third term in our functional
is not conserved along general solutions makes the terminology
``Casimir functional'' questionable, but we stick to it for lack of a
better alternative. The situation is similar to the
one of anisotropic, spherically symmetric steady states 
with $f_0$ depending on the particle energy and the modulus of
the angular momentum, $|x \times v|$, where stability against
spherically symmetric perturbations holds, cf.\ \cite{GR2},
but the general case is open.
For isotropic steady states where $f_0$ depends only on the particle
energy no such restriction is necessary, cf.\ \cite{GR3,R4}.

We want to stress the fact that in this article 
every conclusion is rigorously derived in the mathematical sense. 
Our method stems from our long term research project 
in which both the existence and the stability of galaxy configurations 
has been mathematically investigated. 
Since the Vlasov-Poisson system can be viewed as an infinite 
dimensional Hamiltonian system, even the 
celebrated Antonov stability 
theorems for polytropes are, strictly speaking,
only valid  in the linearized sense. 
This is like the case of center equilibria in a $2     
\times 2$ Hamiltonian system: in general no dynamical stability assertion
can be deduced from a linearized analysis, and some Liapunov type methods are 
needed. 
Recently, we were able to verify the validity of Antonov's 
theorems in a dynamical, nonlinear sense \cite{GR3}. Our variational 
method also yields stability of certain Camm models \cite{GR2}.

The paper proceeds as follows: In the next section we formulate the basic set
up and state our main results. In order to put these into perspective
we include a short discussion
of stability concepts from a mathematics point of view.
In the third section we 
discuss various examples and some further properties
of the steady states which we obtain.
For interested readers we collect 
the details of our mathematical proofs in an appendix. 

We conclude the introduction with some references to the 
mathematical literature.
In addition to our work mentioned above the stability of spherically
symmetric steady states
of the Vlasov-Poisson
system in the present stellar dynamics case is also investigated in
\cite{A,Wa,Wo}. 
Global classical solutions 
to the initial value problem for the Vlasov-Poisson system were first 
established in \cite{P}, cf.\ also \cite{S} and \cite{R7}.  
For the plasma physics case
where the sign in the Poisson equation is
reversed, the stability problem is better
understood, and we refer to \cite{BRV,GS1,GS2,R1}.  
A rather general condition
which guarantees finite mass and compact support of steady
states, but not their stability, is established in \cite{RR}.

\section{Main results}
\setcounter{equation}{0}
Let us first fix some notation.
For a measurable function $f=f(x,v)\ge 0$ we define its induced 
spatial density
\[
\rho_f (x):= \int f(x,v)\, dv,\ x \in \R^3,
\]
and its induced gravitational potential
\[  
U_f :=  - \rho_f \ast \frac{1}{|\cdot|}.
\]
Next we define
\begin{eqnarray*}
\ekin (f)
&:=&
\frac{1}{2} \int\!\!\int |v|^2 f(x,v)\,dv\,dx, \\
\epot (f)
&:=&
- \frac{1}{8\pi} \int |\nabla U_f (x)|^2 dx = 
- \frac{1}{2} \int\!\!\int\frac{\rho_f(x) \rho_f(y)}{|x-y|}dx\,dy,\\
\C(f)
&:=&
\int\!\!\int Q(f(x,v),P(x,v))\,dv\,dx 
\end{eqnarray*}
where 
$Q$ is a given function satisfying certain assumptions specified 
below and $P$ is defined in (\ref{pdef}).
We will minimize the energy-Casimir functional
\[
\HC := \ekin + \epot + \C
\]
under a mass constraint, i.~e., over the set
\[
\F_M := \Bigl\{ f \in L^1(\R^6) 
\mid
f \geq 0,\ \int\!\!\int f(x,v)\,dv\,dx = M,\ \C(f) + \ekin(f) < \infty\Bigr\},
\]
where $M>0$ is prescribed. Eventually, we need to restrict
ourselves to axially symmetric functions in this set:
\begin{eqnarray*}
\F_M^S 
:= \Bigl\{ f \in \F_M 
&\mid&
f(Ax,Av)=f(x,v),\ x,v \in \R^3,\\
&&\ A\ \mbox{any rotation about the $x_3$-axis}\Bigr\}.
\end{eqnarray*}
For the function $Q$ determining the
Casimir functional we make the following  

\smallskip
\noindent {\bf Assumptions on $Q$}: 
$Q,\ \partial_f Q,\ \partial_f^2 Q \in C ([0,\infty[\times \R)$, 
$Q \geq 0$, $Q(0,\cdot)=0=\partial_f Q(0,\cdot)$, and with
constants $C,\ C',\ F,\ F'>0$, and $0 < k,\ k' < 3/2$,
\begin{itemize}
\item[(Q1)]
$Q(f,P) \geq C \,f^{1+1/k},\ f \geq F$, $P\in \R$,
\item[(Q2)]
$\partial_f^2 Q (f,P)> 0,\ f >0,\ P \in \R$,
\item[(Q3)]
$Q(f,\cdot)$ is increasing on $]-\infty,0[$ and decreasing on $]0,\infty[$,
$f \geq 0$,
\item[(Q4)]
$Q(f,0) \leq C' \,f^{1+1/k'},\ f \leq F'$.
\end{itemize}
The first two assumptions are essential whereas the assumptions (Q3)
and (Q4)
can be modified in various ways, cf.\ the next section.
Note that under the assumptions above 
$\partial_f Q(\cdot,P): [0,\infty[ \to [0,\infty[$
is one-to-one and onto for any $P \in \R$.
The steady states which we shall
obtain will be of the form
\[
f_0(x,v)=\phi(E_0-E,P),
\]
where the particle energy $E$ is defined in terms of the induced
potential $U_0$ as in (\ref{edef}),
\begin{equation} \label{phidef}
\phi(E,P):=\left\{
\begin{array}{ccl}
(\partial_f Q (\cdot,P))^{-1}(E)&,&\ E\geq 0,\ P \in \R,\\
0&,&\ E < 0,\ P \in \R,
\end{array}
\right.
\end{equation}
and $E_0\leq 0$ is some cut-off energy.
We should point out that if $Q$ does not depend on $P$---a case which
is included in our assumptions---then no symmetry assumptions need
to be made in the choice of the set $\F_M$ nor anywhere else.
We now state our first main result:

\begin{theorem} \label{propminim}
For every $M>0$ there is a minimizer $f_0 \in \F_M^S$ 
of the energy-Casimir functional $\HC$.
Let $U_0$ denote the potential induced by $f_0$.
Then $f_0$ is a function of the particle energy
$E$ and angular momentum $P$ as defined in (\ref{edef})
and (\ref{pdef}),
\[
f_0 (x,v)=\phi(E_0-E,P)
\]
where $\phi$ is defined in terms of $Q$ by (\ref{phidef}).
The parameter $E_0$ plays the role of a Lagrange multiplier
and is given by
\[
E_0 := \frac{1}{M} 
\int\!\!\int \left[E + \partial_f Q(f_0,P)\right]\,f_0 dv\, dx. 
\]
In particular, $f_0$ is an axially symmetric steady state of the 
Vlasov-Poisson system with total mass $M$.
\end{theorem} 

The crucial part here is the existence assertion
a detailed proof of which is given in the
appendix. If one computes the Euler-Lagrange condition 
corresponding to our variational problem one obtains 
the assertions on the form of the minimizer.
For the details of this tedious but straight forward argument
we refer to \cite{GR2} or \cite{GR3}.

{\bf Remark~1.}
If one is interested only in the existence of a steady state
the usual approach is to prescribe some suitable function
$f_0(x,v)=\phi(E_0-E,P)$, compute the corresponding spatial density 
$\rho_0$ which becomes, via $E$, a functional of the potential $U_0$,
and one is then left with the problem of proving that the semilinear
Poisson equation
\begin{equation} \label{semilinp}
\lap U_0 = 4 \pi \int \phi\left(E_0-\frac{1}{2}|v|^2 + U_0,P(x,v)\right)\, dv
\end{equation}
has a suitable solution.
The non-trivial problem here is to decide
which solutions have finite total mass and compact support in space.
The first requirement
for $\phi$ from the point of view of our theorem is that $\phi$ is a
strictly increasing function of $E_0-E$ so that a corresponding
function $Q$ and the Casimir functional can be defined.
Of course, for the existence question alone this monotonicity
condition which in astrophysics textbooks appears as a stability
condition is not necessary. A mathematical approach which makes no such
monotonicity assumption can be found in \cite{RR}. The point is
that the minimizer which we obtain is not just any old
steady state, but a nonlinearly stable one.

{\bf Some remarks on stability concepts.}
It may be useful to briefly review various
concepts of stability before we state and discuss our stability theorem;
for a more detailed such discussion we refer to \cite{HMRW}.
Consider a nonlinear dynamical system which for short we write as
\[
\dot x = A(x),
\]
$A$ being some nonlinear operator on the possibly infinite dimensional  
state space where solutions $t\mapsto x(t)$ take their values. 
Let $x_0$ be a steady state, i.~e., $A(x_0)=0$. 
\begin{itemize}
\item
The ``appropriate'' stability approach is certainly one where we can deal
with our problem directly, i.~e., without simplifying or linearizing it in
any way. The steady state is called {\em nonlinearly stable}
or {\em Liapunov stable} if one can guarantee that sufficiently small
perturbations of the steady state launch solutions which are
arbitrarily small perturbations of the steady state for future times.
More formally: One must find
two norms $\|\cdot\|,\ \|\!|\cdot|\!\|$ on the state space such that
for any (small) number $\epsilon >0$ there always exists another (small) 
number $\delta >0$ such that every solution $x(t)$ with 
$\|x(0)-x_0\| < \delta$ satisfies 
$\|\!|x(t)-x_0|\!\| < \epsilon$ for all $t\geq 0$.
Some comments are in order: 
\begin{itemize}
\item
It is desirable but not always possible to
use the same norms for measuring the initial perturbation and the
one at time $t$; only on finite dimensional spaces are
all norms  equivalent.
\item
Sometimes one may not even obtain norms but more general tools
to measure the deviations with.
\item
From a physics point of view it is unsatisfactory if no
rule is given on how to obtain the $\delta$ if $\epsilon$ is prescribed. 
From this point of view a better result would be an estimate 
of the form
\[
\|\!|x(t)-x_0|\!\| \leq C\,\|x(0)-x_0\|,\ t \geq 0
\]
possibly with some explicit constant $C$. Unfortunately, mathematicians 
are not always (not often?) able to provide this.
\item
A global existence result which guarantees that solutions exist for all
time $t\geq 0$ at least for initial data close to the steady state is
a necessary prerequisite and integral part for all the above.
\end{itemize}
\item
Since assessing nonlinear 
stability in the sense above is difficult, 
the problem is often approached via linearization. 
By Taylor expansion
one can compute the linearization of $A$ at $x_0$; formally 
$A_{\rm lin} = D_x A(x_0)$.
There are now at least two sub-concepts of {\em linearized stability}:
One is to repeat the definition for Liapunov stability given above,
but replace $A$ by its linearization $A_{\rm lin}$. After all,
saying that $A$ is nonlinear does not forbid $A$ happening to be linear.
This is what we want to call {\em linearized stability}.
A somewhat different concept is {\em spectral stability}:
One considers solutions of the linearized equation of the form
$e^{\lambda t} x$ and tries to find out what the possible eigenvalues
$\lambda$ are. If the real parts of all these eigenvalues are strictly
negative then one calls the steady state {\em spectrally stable}.
Again, some comments are in order:
\begin{itemize}
\item
In general there is no guarantee that the linearized problem 
actually has solutions of the form $e^{\lambda t} x$. If it does,
then in order to draw any conclusions one must know that there are
sufficiently
many such solutions to get the general solution of the linearized problem
by superposition (completeness of the eigenfunctions).
And even if this goes well, there is simply no general result which
will allow any conclusion on the behavior of the original, nonlinear
system, unless the system is finite dimensional!
\item
If one is dealing with a conservative system with some sort of
Hamiltonian structure (such as the Vlasov-Poisson system),
then the best to expect as far as spectral stability is concerned may be
that all the eigenvalues are purely imaginary (due to inherent symmetries
of the spectrum). In this case
no stability follows for the nonlinear system, not even in the
finite dimensional case.
\item
Assume one can establish the existence of a growing mode, i.~e.,
of a solution of the form $e^{\lambda t} x$ where $\lambda$ has positive 
real part. Then again there is no general result saying that
$x_0$ is now also nonlinearly unstable, unless the system is finite 
dimensional. However, such a growing mode is a valuable first step toward
proving a nonlinear instability result. How difficult it still may be
to get to a mathematically rigorous nonlinear instability result
from there is illustrated by \cite{GS1, GS2}
\end{itemize}
\end{itemize}
The upshot from all this is: One should be very careful to draw
conclusions about stability for nonlinear, infinite dimensional 
systems from linearization, in particular, for Hamiltonian ones.

All this said let us now return to our minimizer $f_0$ from 
Theorem~\ref{propminim}.
To investigate its dynamical stability we note first
that if $\int\!\!\int f=M$,
\begin{equation} \label{d-d}
\HC (f)- \HC (f_0)=d(f,f_0)-\frac{1}{8 \pi}
\|\nabla U_f-\nabla U_0\|^2_2,
\end{equation}
where $\|\cdot\|_2$ denotes the usual norm on $L^2(\R^3)$ and
\[
d(f,f_0) := 
\int\!\!\int \Bigl[Q(f,P)-Q(f_0,P) + (E - E_0)(f-f_0)\Bigr]\,dv\,dx;
\]
we are allowed to subtract
the term $E_0 (f-f_0)$ from the integrand since its integral vanishes.
Since $Q(\cdot,P)$ is convex, 
the integrand can be estimated from below by
\[
\Bigl[\partial_f Q(f_0,P)+ E - E_0\Bigr]\,(f-f_0).
\]
According to Theorem~\ref{propminim},
this quantity is zero
where $f_0>0$, while it equals $(E-E_0)\,f \geq 0$
where $f_0=0$.
Thus we see that
\[
d(f,f_0) \geq 0,\ f \in \F_M.
\]
Now we note that the left
hand side of (\ref{d-d}) is constant along axially symmetric solutions.
Moreover, and this is the crucial point,
the term $\|\nabla U_f-\nabla U_0\|^2_2$ which seems to have the
wrong sign in (\ref{d-d}) vanishes along minimizing
sequences by Theorem~A\ref{exminim} stated in the appendix. 

Before we state our stability result we point out
that if we shift a minimizer in the $x_3$ direction we obtain another
minimizer. Moreover, we do in general not know 
whether the minimizers are unique up to spatial shifts.
As discussed in the next section
a minimizer must a-posteriori
have the following additional symmetry property
\[
f_0(x_1,x_2,x_3^\ast-x_3,v_1,v_2,-v_3)
=f_0(x_1,x_2,x_3^\ast+x_3,v_1,v_2,v_3),\ x,v \in\R^3,
\]
for some $x_3^\ast \in \R$. If we take without loss of generality
$x_3^\ast =0$ we refer to this as {\em reflexion symmetry}. 
This symmetry propagates along solutions of the time dependent problem,
and, by restricting ourselves to data with this symmetry, we can
at least ignore the non-uniqueness due to shifts:

\begin{theorem} \label{stability}
For every $\epsilon>0$ there is a $\delta>0$ such that
for any solution $t \mapsto f(t)$ of the Vlasov-Poisson system
with $f(0) \in C^1_c (\R^6) \cap  \F_M^S$ and reflexion symmetric,
\[
d(f(0),f_0) + \frac{1}{8\pi} \|\nabla U_{f(0)}-\nabla U_0\|_2^2
< \delta
\]
implies that for all $t \geq 0$,
\[
d(f(t), f_0) + \frac{1}{8\pi} 
\|\nabla U_{f(t)}-\nabla U_{f_0}\|_2^2 < \epsilon,
\]
{\em provided} $f_0$ is unique or at least isolated in the 
reflexion symmetric subset of $\F_M^S$.
\end{theorem}

Since after what was said above the proof is simple and instructive
we include it here:

\proof
Assume the assertion were false. 
Then there exist $\epsilon>0$, $t_n>0$, and
$f_n(0) \in C^1_c (\R^6) \cap \F_M^S$ reflexion symmetric and such that
for all $n \in \N$, 
\begin{equation} \label{dist0}
d(f_n(0),f_0) + \frac{1}{8\pi} \|\nabla U_{f_n(0)}-\nabla U_0\|_2^2
< \frac{1}{n}
\end{equation}
but
\begin{equation} \label{disttn}
d(f_n (t_n),f_0) + \frac{1}{8\pi} 
\|\nabla U_{f_n(t_n)}-\nabla U_0\|_2^2
\geq \epsilon.
\end{equation}
By (\ref{dist0}) and (\ref{d-d}),
\[
\lim_{n\to \infty}\HC(f_n (0)) = h_M^S,
\]
where $h_M^S$ is the finite, negative minimum of $\HC$ on $\F_M^S$,
cf.\ Lemma~A~\ref{scaling}.
Since $\HC$ is conserved along classical, axially symmetric solutions as
launched by $f_n (0)$, since mass is conserved, and since the
assumed symmetry propagates,
\[
\lim_{n\to \infty}\HC( f_n (t_n)) =  h_M^S \ \mbox{and}\ 
f_n (t_n) \in \F_M^S,\ n \in \N,
\]
i.~e., $(f_n (t_n))$ is a reflexion symmetric
minimizing sequence for $\HC$ in $\F_M^S$.
Up to a subsequence we may therefore assume by Theorem~A~\ref{exminim}
that
\begin{equation} \label{fieldn}
\|\nabla U_{f_n (t_n)}-\nabla U_{f_0}\|^2_2 \to 0;
\end{equation}
note that due to the reflexion symmetry of the minimizing sequence
and any minimizer the shifts along the $x_3$-axis in the statement of
Theorem~A~\ref{exminim} must be zero in the present situation.
It is at this point that the uniqueness or isolation of the minimizer
$f_0$ is used.
Since 
$\lim_{n\to \infty}\HC(f_n (t_n)) = h_M^S = \HC (f_0)$
we conclude by (\ref{fieldn}) and (\ref{d-d}) that
\[
d(f_n (t_n),f_0) \to 0,\ n \to \infty,
\]
and we
arrive at a contradiction to (\ref{disttn}). \prfe

{\bf Remark 2.} If the minimizer is not isolated, 
i.~e., if arbitrarily close to it with respect to
the ``distance'' used in the theorem there are other minimizers
which do not result from a simple translation in space,
then we obtain a stability result of the following form:

Let $\M_M \subset \F_M^S$ denote the set of all minimizers of
$\HC$ in $\F_M^S$. Then
for every $\epsilon>0$ there is a $\delta>0$ such that
for any solution $t \mapsto f(t)$ of the Vlasov-Poisson system
with $f(0) \in C^1_c (\R^6)\cap \F_M^S$,
\[
\inf_{f_0 \in \M_M} 
\left[ 
d(f(0),f_0) + \frac{1}{8\pi} \|\nabla U_{f(0)}-\nabla U_0\|_2^2
\right] < \delta
\]
implies that for all $t\geq 0$,
\[
\inf_{f_0 \in \M_M}
\left[ 
d( f (t),f_0) + \frac{1}{8\pi} 
\|\nabla U_{ f (t)}-\nabla U_{f_0}\|_2^2 \right]< \epsilon.
\]

{\bf Remark 3.}
Although we only showed that $d(f,f_0)\geq 0$ for $f \in \F_M$, one may 
think of this term as a weighted $L^2$-difference of $f$ and $f_0$,
and by a Taylor expansion a stronger and more explicit estimate 
can be obtained if
$\partial_f^2 Q$ is bounded away from zero.

{\bf Remark 4.}
The restriction $f(0) \in \F_M$ for the perturbed initial data would be
acceptable from a physics point of view: A perturbation
of a given galaxy, say by the gravitational pull of some outside object,
results in a perturbed state which is just a rearrangement
of the original state, in particular, its mass remains unchanged.
But in general symmetry properties of the steady state will be
destroyed by such a perturbation.
However, we need that $f(0)$ is axially symmetric so that
$\C$ is conserved along the resulting axially symmetric solution
$f(t)$. So while this symmetry restriction for the perturbation
seems necessary within the present mathematical framework it is 
quite undesirable from a physics point of view.
If the steady state does not depend on $P$ and thus is isotropic,
then no symmetry restrictions are necessary anywhere in our results,
and the stability is with respect to quite general perturbations.

{\bf Remark 5.}
Our stability result is of the nonlinear Liapunov type discussed above in
general, but it suffers from the defect that given $\epsilon$
it does not say how $\delta$ needs to be chosen.
The reason for this is that our
proof is by contradiction and relies on a compactness result. 

\section{Examples and further properties}
\setcounter{equation}{0}

In this section we investigate some additional properties of the resulting
steady states, keeping the discussion somewhat informal.
First we present some examples for functions $Q$ which
satisfy our assumptions or possible variations of them:

{\em Examples.}
A simple class of examples for functions $Q$ which satisfy
the assumptions (Q1)--(Q4) is given by
\begin{equation} \label{ex1}
Q(f,P) = f^{1+1/k} g(P),\ f\geq 0,\ P \in \R,
\end{equation}
where $0<k<3/2$ and $g$ is positive, continuous, bounded, bounded
away from zero, increasing on $]-\infty,0[$, and decreasing on $]0,\infty[$.
Note that in this case,
\[
f_0(x,v)=C (E_0-E)^k g(P)^{-k}
\]
on its support.
Examining the proof of Lemma~A~\ref{scaling} in the appendix, 
which is the only place
where the assumptions (Q3) and (Q4) enter, one can see that these 
assumptions can for example be replaced by
\begin{itemize}
\item[(Q3')]
$Q(\lambda f,P)\geq \lambda^{1+1/k'} Q(f,P),\ f\geq 0,\ 0\leq \lambda \leq 1,\
P\in \R$, with $1/2 \leq k' < 3/2$,
\item[(Q4')]
$Q(f,P) \leq C' \,f^{1+1/k''},\ f \leq F',\ |P|\geq P_0$,  with constants
$C'>0,\ F'>0,\ P_0>0$, and $0 < k'' < 3/2$.
\end{itemize}
Examples which satisfy the assumptions (Q1), (Q2), (Q3'), (Q4')
but not the original ones are provided by (\ref{ex1}) with $1/2 \leq k < 3/2$
and $g$ which has all the properties stated above except
the monotonicity.

{\em Regularity and boundary condition.}
Since by Theorem~\ref{propminim} 
$f_0$ is a function of the quantities $E$ and $P$
which are constant along particle trajectories we are justified
to call $f_0$ a steady state provided $U_0$ is sufficiently
regular to allow for the definition of particle trajectories 
to begin with.
Now by construction $\lap U_0 = 4 \pi \rho_0$ on $\R^3$, at least
in the sense of distributions, and from the very construction
of $f_0$ certain integrability properties of $\rho_0$ follow,
cf.\ the appendix.
Applying the usual Sobolev space arguments one can eventually conclude
that $U_0$ is twice continuously differentiable and 
$\lim_{|x|\to\infty} U_0(x)=0$, cf.\ \cite[Thm.~3]{GR3} for the
technical details.

{\em Finite mass and compact support.}
By construction, the steady states have finite mass $M>0$ which we prescribe
by our mass constraint. 
To continue
we note that by Theorem~\ref{propminim} and a change of variables,
\begin{eqnarray} \label{rhorep}
\rho_0(x)
&=&
\int f_0(x,v)\,dv \nonumber\\
&=&
2 \pi \int_{U_0(x)}^{E_0} \int_{-\sqrt{2(E-U_0(x))}}^{\sqrt{2(E-U_0(x))}}
\phi (E_0-E,r(x)\,p)\,dp\,dE
\end{eqnarray}
where $U_0(x) < E_0$, and $\rho_0$ is zero else; 
here $r(x):=\sqrt{x_1^2 + x_2^2}$.
Assume 
that $\lim_{|x| \to \infty} U_0 (x) = 0$, see the discussion above.
Then clearly the cut-off energy cannot be positive, i.~e., $E_0 \leq 0$,
since otherwise we get infinite mass ``at spatial infinity''.
Since $U_0 < 0$ everywhere, (\ref{rhorep}) shows that $\rho_0$
will have compact support if and only if $E_0 < 0$. An additional, 
sufficient condition which implies this is
\begin{equation} \label{finradcon1}
f\, \partial_f Q(f,P) \leq 3 Q(f,P),\ f\geq 0,\ P \in \R,
\end{equation}
which holds for example for $Q(f,P)=f^{1+1/k} g(P)$ with $1/2 \leq k < 3/2$
and some function $g$. 
To see this we rewrite the formula for $E_0$
from Theorem~\ref{propminim}:
\[
E_0 =
\frac{1}{M} \left[ \ekin(f_0) + 2 \epot(f_0) + 
\int\!\!\int f_0\partial_f Q(f_0,P)\,dv\,dx \right].
\]
Now we note that for solutions of the Vlasov-Poisson system,
\[
\frac{d}{dt} \int\!\!\int x\cdot v f = 2 \ekin(f) + \epot(f)
\]
so that for a steady state the right hand side is zero. Thus by 
(\ref{finradcon1}),
\begin{eqnarray*}
\ekin(f_0) + 2 \epot(f_0) + 
\int\!\!\int f_0\partial_f Q(f_0,P)\,dv\,dx 
&\leq& \\
\ekin(f_0) + 2 \epot(f_0) + 
3 \int\!\!\int Q(f_0,P)\,dv\,dx
&=& \\
3 \ekin(f_0) + 3 \epot(f_0) + 3 \C(f_0)
&=& 3 \HC(f_0) < 0
\end{eqnarray*}
by Lemma~A~\ref{scaling} (a). An alternative condition which also
guarantees compact support is the following:
\begin{equation} \label{finradcon2}
\left.
\begin{array}{c}
\phi(E_0-E,P) \leq C_1 (E_0-E)^{k_1},\ E \to -\infty,\ P\in \R, \\
\phi(E_0-E,P) \geq C_2 (E_0-E)^{k_2},\ E \to E_0 -,\ P \in \R
\end{array}
\right\}
\end{equation}
for positive constants $C_1,\ C_2$ and $0<k_1,k_2<3/2$.
This assumption also implies that $E_0 < 0$, cf.\ \cite[Thm.~3]{R4}.

In the spherically symmetric case our approach provides an explicit 
bound on the radius of the steady state, cf.\ \cite{GR1}.
Note that in this case
one can under appropriate assumptions decide whether
$U_0$ crosses the cut-off energy level $E_0$ by direct examination of
the semilinear Poisson equation (\ref{semilinp}),
since this equation reduces to an ordinary differential equation
with respect to the radial variable, cf.\ \cite{RR}.
However, in the axially symmetric case this is no longer true,
and the corresponding analysis of the genuine partial differential
equation (\ref{semilinp}) would be much more difficult.

{\em Symmetry.}
By construction the steady states which we obtain are
axially symmetric, but so are the spherically symmetric
ones whose existence was known already. In order to make sure
that we have found qualitatively new steady states we
must show that they are in general not spherically
symmetric.
To see this we take $Q$ such that $\phi(E_0-E,\cdot)$
is not constant on any neighborhood of $P=0$. We claim that neither
$\rho_0$ nor $U_0$ are spherically symmetric in this case.
Indeed, if $\rho_0$ were spherically symmetric the same were
true for $U_0$. If $U_0$ were spherically symmetric we take
some $x \in \R^3$ with $r(x)=\sqrt{x_1^2 + x_2^2}\neq 0$ small and
take $A\in {\rm SO}(3)$ such that $Ax=(0,0,|x|)$, hence
$r(Ax)=0\neq r(x)$ but $U_0(Ax)=U_0(x)$. Inserting this into
the formula (\ref{rhorep}) for $\rho_0$ we see that in general
$\rho_0(Ax)\neq \rho_0(x)$ so $\rho_0$ is not spherically symmetric.

Indeed, this is a bit more than just saying that $f_0$ is not
spherically symmetric. To see this, take a spherically symmetric
steady state $f_0$ with induced density $\rho_0$, potential $U_0$,
and particle energy $E=|v|^2/2 +U_0(x)$.
Let $g_0(x,v)=\psi(E,P)$  
be any function which is odd in $P$,
and such that $f_0+g_0\geq 0$. Since $P$ is not invariant under general
rotations neither is $f_0+g_0$, but the fact that $\psi$ is odd in $P$
implies that $\rho_{f_0+g_0}=\rho_0$ and the same holds for the potential.
So this trivial construction gives an axially symmetric steady state
where the phase space density is not spherically symmetric but
the macroscopic quantities $\rho_0$ and $U_0$ are. Our steady states
are in general not of this trivial type.

Another symmetry issue refers to the dependence of the minimizer
on the variable $x_3$. Given a minimizer $f_0$ of which we do not yet
know any symmetry with respect to $x_3$ we can
do a symmetric decreasing rearrangement of $f_0$
with respect to $x_3$ while keeping all other variables fixed.
Denote this rearrangement by $f_0^\ast$; as to the rearrangement concept we
refer the reader to \cite[Ch.~3]{LL}. Obviously, the rearrangement does not 
change the kinetic energy, and neither does it change the Casimir functional,
since $P$ does not depend on $x_3$. By \cite[Thms.~3.7, 3.9]{LL}
it can at most decrease the potential energy, with equality iff
$f_0$ is already symmetric and decreasing in $x_3$ up to a shift in $x_3$.
But since $f_0$ minimizes the energy-Casimir functional it must
posses this symmetry, and without loss of generality we can assume
that $f_0$ is reflexion symmetric in $x_3$. The corresponding symmetry
in the variable $v_3$ follows from the form which $f_0$ must have
due to Theorem~\ref{propminim}.

{\em Stationary versus static solutions.}
If instead of the Vlasov-Poisson system we consider a self-gravitating fluid
as described by the Euler-Poisson system then
every so-called static solution, i.~e., every steady state with
vanishing velocity field, is spherically symmetric, cf.\ \cite{Li}. 
This turns out to be false for the steady
states of the Vlasov-Poisson system which we obtain.
By definition the velocity field equals $j_0/\rho_0$ on the support
of $\rho_0$ where 
the mass current density $j_0$ is given by
\[
j_0(x) =
\int v\,f_0(x,v)\, dv =
2 \pi \int_{U_0(x)}^{E_0} \int_{-\sqrt{2(E-U_0(x))}}^{\sqrt{2(E-U_0(x))}}
s\, \phi(E_0-E,r(x)\,p)\,dp\,dE \; \; e_t(x)
\]
with $e_t(x):=(-x_2,x_1,0)/r(x)$; on the $x_3$-axis the velocity field
vanishes.
Obviously, if $\phi$ is even in $P$ then $j_0$ vanishes
identically,
so there exist static solutions which are not spherically symmetric
among the ones we obtain. On the other hand, if, say,
$\phi(E,P) > \phi(E,-P)$ for all
$E\geq 0$ and $P>0$ the velocity field is non-trivial
and corresponds to an average rotation about the axis of symmetry in the
counterclockwise direction. 

{\em Less regular dependence on $P$}.
When examining the appendix the reader may notice 
that at no place we really make use of the fact
that all the functions under consideration can be taken to
be axially symmetric. This is because as opposed to
spherical symmetry, exploiting axial symmetry for example in the estimates
for the potential is technically quite unpleasant. However, it is possible
and even necessary if one wishes to study examples of the form
\begin{equation} \label{poly}
Q(f,P) = f^{1+1/k} P^{-2l/k},\ f \geq 0,\ P \in \R
\end{equation}
which for $l\neq 0$ do not satisfy the assumptions on $Q$ which 
we stated above.
The corresponding steady states would take the form
\[
f_0(x,v)=C (E_0-E)^k P^{2l}
\]
which is analogous to the classical, spherically symmetric polytropes,
except that the square of the third component of the angular momentum
replaces the square of the modulus of the angular momentum. 
Clearly, (\ref{poly}) satisfies the crucial convexity condition (Q2),
provided $k>0$. When we examine the scaling arguments in Lemma~A~\ref{scaling}
we find that they go through, provided
\[
l>-1,\ 0<k<l+\frac{3}{2}.
\]
The place where significant changes become necessary are
the estimates in Lemma~A~\ref{rhoest} and Lemma~A~\ref{potcomp}.
Clearly, one can no longer control $\|f\|_{1+1/k}$ and $\|\rho_f\|_{1+1/n}$
in terms of $\ekin(f)$, $\C(f)$, and $M$. Instead, one gains control
of the weighted norms
\[
\int\!\!\int P^{-l/k}(x,v)\,f^{1+1/k}(x,v)\, dv\,dx,\qquad
\int r^{-2l/n}(x)\,\rho_f^{1+1/n}(x)\,dx
\]
where $r(x)=\sqrt{x_1^2+x_2^2}$. We conjecture that the weight factor
can be dealt with in the estimates for the potential energy etc.\
if one makes proper use of the axial symmetry of the potential.

\section{Appendix}
\setcounter{equation}{0}
We now give the proof of the existence part in Theorem~\ref{propminim}
and the additional results used in the proof of Theorem~\ref{stability}. 
First we collect some estimates
for $\rho_f$ and $U_f$ induced by an element $f \in \F_M$.
These estimates make no use of symmetry and rely on the
assumption (Q1). Their main point is to see that
$\HC$ is bounded from below on $\F_M$ and to establish
certain bounds along minimizing sequences of $\HC$.
Constants
denoted by $C$ are positive, may depend only on $Q$ and $M$, and 
their value may change from line to line. By $\|\cdot\|_p$
we denote the usual $L^p$-norm of functions over $\R^3$
or $\R^6$ as the case may be.

\begin{lemmaa} \label{rhoest}
Let $n := k+3/2<3$. Then for any $f \in \F_M$ the following holds:
\begin{itemize}
\item[{\rm (a)}]
$ \displaystyle \hspace{2.9cm}
\int\!\!\int f^{1+1/k}(x,v)\, dv\,dx \leq C (1 + \C(f))
$
\item[{\rm (b)}]
$ \displaystyle \hspace{2.5cm}
\int \rho_f^{1+1/n}(x)\,dx 
\leq
 C\, \left(1 + \ekin(f) + \C(f)\right)
$
\item[{\rm (c)}]
$U_f \in L^{6}(\R^3)$ with $\nabla U_f \in L^{2}(\R^3)$, the two forms
of $\epot (f)$ stated in Section~2 are equal, and
\[
\int |\nabla U_f|^2 dx \leq
C\,\|\rho_f\|_{6/5}^2
\leq C \, \left(1 + \ekin (f) + \C(f)\right)^{n/3}.
\]
\end{itemize}
\end{lemmaa}

\proof  Part (a) is a direct consequence
of (Q1), splitting the
$v$-integral in the definition of $\rho_f$
into small and large $v$'s and optimizing the split yields (b),
while the extended Young's inequality and interpolation
together with (b) imply (c).
For details cf.\ \cite{GR2} or \cite{GR3}. \prfe

The estimates above have an immediate
but important consequence:

\begin{lemmaa} \label{minseqbounds}
The energy-Casimir functional $\HC$ is bounded from below on
$\F_M$, i.~e.,
\[
h_M := \inf_{f\in \F_M} \HC(f) > - \infty,
\]
and along any minimizing sequence of $\HC$ in $\F_M$
the quantities $\ekin(f)$, $\C(f)$,
$\|f\|_{1+1/k}$, and $\|\rho_f\|_{1+1/n}$
are bounded.
\end{lemmaa}

\proof
Lemma~A~\ref{rhoest} yields the estimate
\[
\HC(f) \geq \ekin(f) + \C(f) - 
C \, \left(1 + \ekin (f) + \C(f)\right)^{n/3},\ f \in \F_M,
\]
and since $n<3$ the assertions follow. \prfe 

Lemma~A~\ref{rhoest} and Lemma~A~\ref{minseqbounds} remain valid
if we replace $\F_M$ by its axially symmetric subset $\F_M^S$.
Of course the infimum of $\HC$ on this smaller set may be
larger; we denote
\[
h_M^S := \inf_{f\in \F_M^S} \HC(f).
\]

The assumptions (Q3) and (Q4) determine the
behavior of $\HC$ under scaling transformations which we 
use to show that $h_M$ is negative and to
relate the $h_M$'s for different values of $M$:

\begin{lemmaa} \label{scaling}
\begin{itemize}
\item[{\rm (a)}]
Let $M>0$. Then $-\infty < h_M  < 0$.
\item[{\rm (b)}]
For $0 < \bar M \leq M$,
\[
h_{\bar M} \geq \left( \bar M/M\right )^{5/3} h_{M}.
\]
\end{itemize}
\end{lemmaa}

\proof 
Given any function $f$, we define a rescaled function 
$\bar f(x,v)=a f(bx,cv)$, where $a,\,b,\,c >0$. Then
\begin{equation} \label{massscale}
\int\!\!\int \bar f \,dv\,dx =
a (b\,c)^{-3} \int\!\!\int f \,dv\,dx, 
\end{equation}
i.~e.\ $f \in \F_M$ iff $\bar f \in \F_{\bar M}$ where
$\bar M = a (b\, c)^{-3} M$. Next
\begin{eqnarray} \label{encasscale}
\HC (\bar f) 
&=&
a\,b^{-3} c^{-5} \ekin(f) + a^2 b^{-5} c^{-6} \epot(f) \nonumber \\ 
&&
{} + (b\,c)^{-3} \int\!\!\int Q(a f(x,v),(b\,c)^{-1}P(x,v))\,dv\,dx.
\end{eqnarray}
To prove (a) we fix a function $f \in \F_1$ 
with $f \leq F'$ and
let $a=M\,(b\,c)^3$ so that $\bar f \in \F_M$.
Then by (Q3) and (Q4) and with positive constants
$C_1, C_2, C_3$ which depend on $f$,
\begin{eqnarray*}
\HC (\bar f) 
&\leq&
M\,c^{-2} \ekin (f) + M^2 b\, \epot (f) + 
M a^{-1} \int\!\!\int Q(a f(x,v),0)\,dv\,dx \\
&\leq&
C_1 c^{-2} - C_2 b + C_3 a^{1/k'},
\end{eqnarray*}
provided $a \leq 1$ so that (Q4) can be applied; note that $\epot(f) < 0$.
We want the negative term to dominate
as $b \to 0$, so we let $c=b^{-\gamma/2}$
for some $\gamma >0$. Then $a=M\, b^{3(1-\gamma/2)}$, and
\[
\HC(\bar f) \leq
C_1 b^\gamma -  C_2 b + C_3 M^{1/k'} b^{3(1-\gamma/2)/k'}.
\]
Since $0 < k' < 3/2$ we can fix $\gamma \in ]1,2[$ 
such that $3 (1-\gamma/2)/k' >1$.
For $b>0$ sufficiently small $\HC(\bar f)$ will then be negative 
and $a=M b^{3 (1-\gamma/2)} <1$ as required. 
This proves part (a) of the lemma.
To prove (b) we choose $a=c=1$ and $b=(\bar M/M)^{-1/3}$ 
so that the mapping 
$\F_M\to \F_{\bar M}$, $f \mapsto \bar f$
is one-to-one and onto and $b^{-1}\leq 1$. By (\ref{encasscale}),
\begin{eqnarray*}
\HC(\bar f)
&=&
b^{-3} \ekin(f) + b^{-5} \epot(f) + 
b^{-3} \int\!\!\int Q(f(x,v),b^{-1} P(x,v))\,dv\,dx\\
&\geq&
b^{-5} \ekin(f) + b^{-5} \epot(f) + 
b^{-5} \int\!\!\int Q(f(x,v),P(x,v))\,dv\,dx\\
&=& 
b^{-5} \HC(f);
\end{eqnarray*}
we multiplied the two positive terms by $b^{-2} \leq 1$ 
and used the monotonicity of $Q$ in the $P$-variable which we required in (Q3).
By the choice of $b$ and the definitions of $h_M$ and $h_{\bar M}$ 
the proof is complete.
\prfe 

It is again obvious that the assertions of the lemma remain valid
if we restrict ourselves to the axially symmetric functions in
$\F_M$. The reader might also check that the scaling arguments
work under the assumptions (Q3') and (Q4') from Section~3 as well.

Next we provide a splitting estimate which will be used to
show that along a minimizing
sequence the mass cannot escape to infinity;
here and in the following we denote for $0<R<S\leq \infty$,
\begin{eqnarray*}
B_R 
&:=& \{x \in \R^3 | |x| \leq R\},\\ 
B_{R,S} 
&:=& 
\{ x \in \R^3 | R \leq |x| < S \}.
\end{eqnarray*}

\begin{lemmaa} \label{split1}
Let $f \in \F_M $. Then
\[
\sup_{a\in\R^3} \int_{a+B_R} f (x,v)\,dv\, dx \geq 
\frac{1}{R M} 
\left( - 2 \epot(f) - \frac{M^2}{R} - 
\frac{C \|\rho_f\|_{1+1/n}}{R^{(5-n)/(n+1)}} \right),\ R>1.
\]
\end{lemmaa}

The proof follows from splitting the potential energy into three
parts  
according to $|x-y|\leq 1/R$, $1/R < |x-y| <  R$, and $R \geq |x-y|$;
for the details we refer to \cite{GR3} or \cite[Lemma~3]{R6}
The splitting estimate above has the following important
consequence for minimizing sequences:

\begin{lemmaa} \label{novanishing}
Let $(f_i)\subset \F_M$ be a minimizing sequence
of $\HC$. Then there exist $\delta_0>0$, $R_0>0$, $i_0 \in \N$,
and a sequence of shift vectors $(a_i)\subset \R^3$
such that 
\[
\int_{a_i+B_R}\int f_i (x,v)\,dv\, dx \geq \delta_0,
\ i \geq i_0,\ R\geq R_0.
\]
If $(f_i)\subset \F_M^S$, i.~e., the minimizing sequence
consists of axially symmetric functions then one can choose
$a_i = (0,0,z_i)$ with $z_i \in \R$, i.~e., only shifts
along the axis of symmetry need to be admitted.
\end{lemmaa}

\proof
By Lemma~A~\ref{minseqbounds},
$(\|\rho_{f_i}\|_{1+1/n})$ is bounded.
By Lemma~A~\ref{scaling} (a) we have
\[
\epot (f_i) \leq \HC (f_i) \leq 
\frac{1}{2} h_M < 0,\ i \geq i_0,
\]
for a suitable $i_0 \in \N$.
Thus by Lemma~A~\ref{split1} there exist $\delta_0 >0$, $R_0>0$,
and a sequence of shift vectors $(a_i) \subset \R^3$ as required.

Now assume that the functions $f_i$ are axially symmetric,
and let $a_i=(\bar a_i,z_i)$ with $\bar a_i \in \R^2$ and $z_i \in \R$.
Suppose we can show that the sequence $(\bar a_i)$ is bounded.
Then we can pick some $\bar R > 0$ such that
\[
(\bar a_i,0) + B_{R_0}  \subset B_{\bar R},\ i \in \N
\]
and thus
\[
a_i + B_{R_0}  \subset (0,0,z_i) + B_{\bar R},\ i \in \N,
\]
which implies our assertion in the symmetric case: For $R \geq \bar R$,
\[
\int_{(0,0,z_i) + B_R}\int f_i (x,v)\,dv\, dx \geq 
\int_{a_i+B_{R_0}}\int f_i (x,v)\,dv\, dx \geq \delta_0,
\ i \geq i_0.
\]
So it suffices to show that $(\bar a_i)$ is bounded. Assume the contrary
and choose some integer $N \in \N$ such that $N\, \delta_0 > M$.
If $|\bar a_i|$ is sufficiently large
then simple geometry tells us that there exist rotations
$A_1,\ldots, A_N$ about the $x_3$-axis such that the balls
$A_k a_i + B_{R_0},\ k=1,\ldots,N$, are pairwise disjoint.
By axial symmetry of $f_i$,
\[
\int_{A_k a_i + B_{R_0}}\int f_i (x,v)\,dv\, dx = 
\int_{a_i+B_{R_0}}\int f_i (x,v)\,dv\, dx \geq \delta_0,\ k=1,\ldots,N,
\]
and since the balls on the left hand side are pairwise disjoint,
\[
\int\!\!\int f_i (x,v)\,dv\, dx \geq 
\sum_{k=1}^N \int_{A_k a_i+ B_{R_0}}\int f_i (x,v)\,dv\, dx \geq
N\, \delta_0 > M
\]
which violates the mass constraint and gives the desired contradiction.
\prfe

We will also need to exploit the well known compactness properties
of the solution operator of the Poisson equation:

\begin{lemmaa} \label{potcomp}
Let $(\rho_i) \subset L^{1+1/n} (\R^3)$ be bounded and
\[
\rho_i \rightharpoonup \rho_0 \ \mbox{weakly in}\ L^{1+1/n} (\R^3) .
\]
\begin{itemize}
\item[{\rm (a)}]
For any $R>0$,  
\[
\nabla U_{{\bf 1}_{B_R} \rho_i} \to \nabla U_{{\bf 1}_{B_R} \rho_0} 
\ \mbox{strongly in}\ L^2 (\R^3).
\]
\item[{\rm (b)}]
If in addition $(\rho_i)$ is bounded in $L^1(\R^3)$, $\rho_0 \in \L^1(\R^3)$, 
and
for any $\epsilon >0$ there exists $R>0$ and $i_0 \in \N$
such that
\[
\int_{|x| \geq R} |\rho_i(x)| \, dx < \epsilon,\ i \geq i_0
\]
then
\[
\nabla U_{\rho_i} \to \nabla U_{\rho_0} 
\ \mbox{strongly in}\ L^2 (\R^3).
\]
\end{itemize}
\end{lemmaa}

For the details of the proof we refer to \cite{GR3}
or \cite[Lemma~5]{R6};
it relies on the fact that over bounded sets the solution
operator to the Poisson equation is compact in the
appropriate spaces and the fact that in the context
of the lemma we can make the contribution to the
potential energy coming from outside sufficiently
large balls small.

We are now ready to prove the existence of a minimizer,
more precisely we prove:

\begin{theorema} \label{exminim}
Let $(f_i)\subset \F_M^S$ be a minimizing sequence 
of $\HC$. Then there exists a sequence of shifts 
$(z_i)\subset \R$ and a subsequence, again denoted
by $(f_i)$, such that for any $\epsilon >0$ there exists $R>0$
with
\[
\int_{(0,0,z_i) + B_R} f_i (x,v)\, dv\, dx \geq M - \epsilon,\ i \in \N,
\]
\[
T f_i := f_i(\cdot + (0,0,z_i),\cdot) \rightharpoonup f_0
\ \mbox{weakly in}\ L^{1+1/k}(\R^6),\ i \to \infty,
\]
and
\[
\int_{B_R} f_0 \geq M - \epsilon .
\]
Finally,
\[
\nabla U_{Tf_i} \to \nabla U_0\ 
\mbox{strongly in}\ L^2(\R^3),\ i \to \infty,
\] 
and $f_0 \in \F_M^S$ is a minimizer of $\HC$.
\end{theorema}

\proof
We split $f \in \F_M^S$ into three different parts:
\[
f = {\bf 1}_{B_{R_1}\times \R^3} f + {\bf 1}_{B_{R_1,R_2}\times \R^3} f + 
{\bf 1}_{B_{R_2,\infty}\times \R^3} f
=: f_1 + f_2 + f_3;
\]
the parameters $R_1 < R_2$ of the split are yet to be determined. The induced
spacial densities are denoted by $\rho_k$, their masses by $M_k$,
and the induced potentials by $U_k$, $k=1,2,3$. With
\[
I_{lm}:= \int\!\!\int\frac{\rho_l (x)\, \rho_m (y)}{|x-y|},\
l, m = 1,2,3,
\]
we have
\[
\HC(f) = \HC(f_1) + \HC(f_2) + \HC(f_3)
- I_{12} - I_{13} - I_{23} .
\]
If we choose $R_2 > 2 R_1$
then
\[
I_{13} \leq \frac{C}{R_2} .
\]
Next, we use the Cauchy-Schwarz inequality, the extended Young's
inequality, and interpolation to get
\begin{eqnarray*}
I_{12} + I_{23}
&=&
\frac{1}{4\pi}
\left|\int \nabla(U_1+U_3)\cdot \nabla U_2 dx \right| 
\leq
C \|\rho_1 + \rho_3\|_{6/5} \|\nabla U_2\|_2 \\
&\leq&
C \|\rho\|_{1+1/n}^{(n+1)/6} \, \|\nabla U_2\|_2.
\end{eqnarray*}   
Using the estimates above and Lemma~A~\ref{scaling} (b)
we find
\begin{eqnarray} \label{split2}
h_M^S - \HC(f)
&\leq&
\left(1-\left(\frac{M_1}{M}\right)^{5/3}
- \left(\frac{M_2}{M}\right)^{5/3}
- \left(\frac{M_3}{M}\right)^{5/3}\right) \, h_M^S 
\nonumber \\
&&
{} + C\, \left(R_2^{-1} + 
\|\rho\|_{1+1/n}^{(n+1)/6} \, \|\nabla U_2\|_2\right) 
\nonumber \\ 
&\leq&
\frac{C}{M^2}\left(M_1 M_2 + M_1 M_3 + M_2 M_3\right)\, h_M^S
\nonumber \\
&&
{} + C \left(R_2^{-1} + 
\|\rho\|_{1+1/n}^{(n+1)/6} \, \|\nabla U_2\|_2\right)
\nonumber \\
&\leq&
C h_M^S \, M_1\, M_3 + 
C\, \left(R_2^{-1} + \|\rho\|_{1+1/n}^{(n+1)/6} \, 
\|\nabla U_2\|_2 \right);
\end{eqnarray}
observe that by Lemma~A~\ref{scaling} (a) $h_M^S < 0$
and that constants denoted by $C$ are positive and depend
on $M$ and $Q$, but not on $R_1$ or $R_2$. 
We want to use (\ref{split2}) to show that up to a 
subsequence and a shift $M_{3}$ becomes small along 
any minimizing sequence  
for $i$ large provided the splitting parameters are suitably chosen. 

So let $(f_i)\subset \F_M^S$ be a minimizing sequence and define
$R_0$, $\delta_0$, and $(z_i)$ according to Lemma~A~\ref{novanishing}.
The shifted sequence $(Tf_i)$ is again minimizing
so by Lemma~A~\ref{minseqbounds} we can pick a subsequence,
again denoted by $(f_i)$,
with a weak limit as stated in the theorem.
Now choose $R_1 > R_0$ so that by Lemma~A~\ref{novanishing},
$M_{i,1} \geq \delta_0$ for $i$ large.  
By (\ref{split2}),
\begin{equation} \label{split3}
-C\, h_M^S \delta_0 M_{i,3} 
\leq
\frac{C}{R_2} + C\, \|\nabla U_{0,2}\|_2 
+ C \|\nabla U_{i,2} - \nabla U_{0,2}\|_2 
+ \HC(Tf_i) - h_M^S
\end{equation}
where $U_{i,l}$ is the potential induced by $f_{i,l}$
which in turn has mass $M_{i,l}$, $i\in \N \cup\{0\}$,
and the index $l=1,2,3$ refers to the splitting. 
Given any
$\epsilon >0$ we increase $R_1 >R_0$ such that
the second term on the right hand side of (\ref{split3})
is small, say less than $\epsilon/4$. Next choose
$R_2 > 2 R_1$ such that the first term is small.
Now that $R_1$ and $R_2$ are fixed, the third term in (\ref{split3}) 
converges to zero by Lemma~A~\ref{potcomp} (a).
Since $(Tf_i)$ is minimizing the remainder in (\ref{split3})
follows suit.
Therefore, for $i$ sufficiently large,
\begin{equation} \label{concentr}
\int_{(0,0,z_i) + B_{R_2}}\int Tf_i = M - M_{i,3}
\geq M - (-C\, h_M^S \delta_0)^{-1} \epsilon .
\end{equation}
Clearly, $f_0 \geq 0$ a.~e., and $f_0$ is axially symmetric.
By weak convergence we have that for any $\epsilon >0$
there exists $R>0$ such that
\[
M \geq \int_{B_R} \int f_0\, dv\, dx \geq M-\epsilon
\]
which in particular implies that $f_0 \in L^1(\R^6)$
with $\int f_0 \,dv\,dx = M$. 
The functional $\C$ is convex,
so by Mazur's Lemma and Fatou's Lemma
\[
\C (f_0)\leq 
\limsup_{i\to \infty} \C (T f_i).
\]
The strong convergence
of the gravitational fields now follows by Lemma~A~\ref{potcomp} (b),
and in particular,
\[
\HC (f_0) \leq \limsup_{i\to \infty} \HC(T f_i) = h_M^S
\]
so that $f_0\in \F_M^S$ is a minimizer of $\HC$.  
\prfe

{\em Acknowledgment:}
The research of YG is supported in part by an NSF grant and a 
Sloan Fellowship. 
GR acknowledges support by the Wittgenstein 
2000 Award of P.~A.~Markowich.

\end{document}